\documentclass{aa}  

\usepackage{graphicx}
\usepackage{xcolor}
\usepackage{txfonts}

\newcommand{\orcid}[1]{\href{https://orcid.org/#1}{\includegraphics[width=10pt]{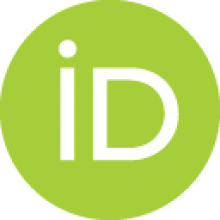}}}
\usepackage{natbib}
\usepackage{multirow}
\usepackage{hyperref}
\hypersetup{linkcolor=red,citecolor=green,filecolor=cyan,urlcolor=magenta}
\hypersetup{
    colorlinks=true,
    linkcolor=blue,
    filecolor=magenta,
    anchorcolor=red,      
    urlcolor=purple,
    citecolor=blue,
    }

\begin{document} 

   \title{The VISCACHA survey}

   \subtitle{XIII. The extended main-sequence turn-off in intermediate-age low-mass clusters}

   \author{S. O. Souza\inst{1,2}\orcid{0000-0001-8052-969X}
           \and
           A. Pérez-Villegas\inst{3}\orcid{0000-0002-5974-3998}
           \and
           B. Dias\inst{4}\orcid{0000-0003-4254-7111}
           \and
           L. O. Kerber\inst{5}\orcid{0000-0002-7435-8748}
           \and
           B. Barbuy\inst{2}\orcid{0000-0001-9264-4417}
           \and 
           R. A. P. Oliveira\inst{6}\orcid{0000-0002-4778-9243}
           \and
           B. P. L. Ferreira\inst{7}\orcid{0000-0002-7552-3063}
           \and
           J. F. C. Santos Jr.\inst{7}\orcid{0000-0003-1794-6356}
           \and
           F. F. S. Maia\inst{8}\orcid{0000-0002-2569-4032}
           \and
           E. Bica\inst{9}\orcid{0000-0003-3336-0910}
           \and
           G. Baume\inst{10,11}\orcid{0000-0002-0114-0502}
           \and
           D. Minniti\inst{4,12}\orcid{0000-0002-7064-099X}
           \and
           E. R. Garro\inst{13}\orcid{0000-0002-4014-1591}
           \and
           A. L. Figueiredo\inst{2}\orcid{0009-0007-1625-8937}
           \and
           J. G. Fernández-Trincado\inst{14}\orcid{0000-0003-3526-5052}
           \and
           S. Saroon\inst{4}\orcid{0000-0002-2789-0934}
           \and
           L. Fraga\inst{15}\orcid{0000-0003-0680-1979}
           \and
           B. Quint\inst{16}\orcid{0000-0002-1557-3560}
           \and
           D. Sanmartim\inst{16}\orcid{0000-0002-9238-9521}
          }

   \institute{Max Planck Institute for Astronomy, K\"onigstuhl 17, D-69117 Heidelberg, Germany -- \email{s-souza@mpia.de}
        \and 
             Universidade de S\~ao Paulo, IAG, Rua do Mat\~ao 1226, Cidade Universit\'aria, S\~ao Paulo 05508-900, Brazil
         \and 
             Instituto de Astronom\'ia, Universidad Nacional Aut\'onoma de M\'exico, A. P. 106, C.P. 22800, Ensenada, B. C., M\'exico
        \and 
             Instituto de Astrofísica, Departamento de Física y Astronomía, Facultad de Ciencias Exactas, Universidad Andres Bello, Fernandez Concha, 700, Las Condes, Santiago, Chile
        \and 
            Universidade Estadual de Santa Cruz, DCEX, Rod. Jorge Amado km 16, Ilhéus 45662-900, Bahia, Brazil
        \and
            Astronomical Observatory, University of Warsaw, Al. Ujazdowskie 4, 00-478 Warszawa, Poland
        \and 
            Departamento de Física, ICEx - UFMG, Av. Antônio Carlos 6627, Belo Horizonte 31270-901, Brazil
        \and 
            Universidade Federal do Rio de Janeiro, Av. Athos da Silveira, 149, Cidade Universitária, Rio de Janeiro 21941-909, Brazil
        \and 
            Universidade Federal do Rio Grande do Sul, Departamento de Astronomia, CP15051, Porto Alegre,91501-970, Brazil
        \and 
            Instituto de Astrofísica de La Plata (CONICET-UNLP), Paseo del bosque S/N, La Plata (B1900FWA), Argentina
        \and 
            Facultad de Ciencias Astronómicas y Geofísicas - Universidad Nacional de La Plata, Paseo del Bosque S/N, La Plata (B1900FWA), Argentina
        \and 
            Vatican Observatory, V00120 Vatican City State, Italy
        \and
            ESO - European Southern Observatory, Alonso de Cordova 3107, Vitacura, Santiago, Chile
        \and 
           Instituto de Astronom\'ia, Universidad Cat\'olica del Norte, Av. Angamos 0610, Antofagasta, Chile
        \and
            Laborat\'{o}rio Nacional de Astrof\'{i}sica LNA/MCTI, 37504-364, Itajub\'{a}, MG, Brazil
        \and
            NSF NOIRLab/NSF–DOE Vera C. Rubin Observatory HQ, 950 N. Cherry Ave., Tucson, AZ 85719, USA (AURA Staff)
             }

   \date{Received ...; accepted ...}

 
  \abstract{
The extended main-sequence turn-off (eMSTO) is a well-known feature observed in young and intermediate-age star clusters, characterized by a significant broadening of the main-sequence turn-off region. Although prolonged star formation and stellar rotation have been proposed as possible explanations, no consensus has yet been reached. Most previous studies have focused on high-mass clusters. In this work, we extend the analysis to the less-explored low-mass regime by investigating star clusters in the Magellanic Clouds using data from the VISCACHA survey. We employed a widely used method to quantify the MSTO width in terms of age spread. Additionally, to validate our approach, we used a cluster also observed with HST. Our analysis confirms that the eMSTO phenomenon is also present in low-mass clusters, following the known {age/mass-MSTO width relations}. In particular, the less massive cluster in our sample does not show an eMSTO, supporting the proposed link between the eMSTO and the escape velocity of the cluster, providing a new lower limit to the age spread of $88\pm40$ Myr for the presence of the eMSTO. The consistent MSTO width measurements between the VISCACHA and HST photometries confirm the robustness of our method and demonstrate that the age spread determination is independent of the photometric system, showing also the power of ground-based observations to investigate the eMSTO phenomenon.
}

\titlerunning{eMSTO in low mass Magellanic Clouds}
\authorrunning{Souza et al.}

\maketitle

\section{Introduction}




Extended main-sequence turn-off (eMSTO) is a common feature in the color-magnitude diagrams (CMDs) of intermediate-age star clusters ($\sim 1–2$ Gyr) first detected in massive clusters of the Magellanic Clouds (MCs) observed with HST \citep{mackey2007,mackey2008,glatt08} and later in Milky Way open clusters (MWOCs) observed with Gaia \citep[][among others]{marino18,bastian2018,cordoni18}. The groundbreaking work by \cite{mackey2007}, followed by many studies \citep[e.g. ][]{goudfrooij2009,goudfrooij2011}, initially interpreted it as evidence of an age spread of 100–700 Myr. This scenario was contested by the discovery of eMSTO in young clusters ($<1$ Gyr) where large age spreads caused by an ongoing star formation history have not been observed so far \citep[e.g.][]{bastian2013,niederhofer15,cabrera-ziri16b}. Alternative interpretations arose, including stellar rotation and binary effects \citep[e.g.][]{bastian2009,yang13, sun19,ettorre25}, but no consensus has been reached.

The first analysis of the star cluster NGC1846 of the Large Magellanic Cloud (LMC) by \cite{mackey2007} using HST/ACS data and the subsequent analysis of a larger sample by \cite{goudfrooij2011,goudfrooij2009,goudfrooij14}, was mainly focused on the relatively high mass stars clusters ($> 2\times 10^4 M_\odot$) with ages spanning from 1 to 2 Gyr. {This mass regime comprehends clusters with more populated CMDs observed with HST}. \citet{goudfrooij14} demonstrated that the MSTO width correlates with the cluster mass, escape velocity, and age, suggesting that the cluster dynamics may influence the eMSTO phenomenon. \cite{niederhofer15} extended these previous results for young massive stars clusters (YMCs), showing, however, that the cluster age also plays an important role for the presence of eMSTO since the YMCs analysed seem not to present this feature in their CMDs. Later, \cite{piatti16aea} analysed four LMC clusters with masses $<5\times10^3M_\odot$, which is approximately one order of magnitude lower than those observed with the HST. They observed that low-mass clusters also have an eMSTO, suggesting that the cluster mass is not the main factor in controlling the phenomenon. 

On the other hand, stellar rotation models have gained traction by reproducing eMSTO morphologies without invoking age spreads, which means that different rotational velocities within a single-age population could broaden the MSTO \citep[e.g.][]{brandt15, georgy19}. One primary effect of stellar rotation is that it carries hydrogen (H) from the outer layers to the nucleus, and this mixing keeps the star in the main-sequence (MS) phase for longer time. {Since different stars have different rotation rates, this effect will also impact them differently, making the MSTO in the CMD wider similarly as an extended star formation} \citep[gyrochronology,][]{bastian2009,vidotto14,niederhofer15}. These models predict a peak in MSTO width (FWHM) at intermediate ages, as stellar rotation effects maximise near the end of the MS, followed by a decline due to angular momentum loss from magnetic braking \citep[e.g.][]{georgy19}. \cite{Bodensteiner23} analysed a sample of Be stars of the $\sim35$Myr old star cluster NGC330 of the Small Magellanic Cloud (SMC), combining photometric data from the HST with spectroscopic data from MUSE. They found that the fast-rotating stars are mainly redder and fainter than the rest of the MSTO. The rotation distribution along the MSTO happens {not only, but also } because the equatorial disc present in some of those stars \citep[e.g. Be;][]{labadie22}, depending on the inclination angle of the disc with respect to the observer, can obscure the star in optical, changing their position in the CMD and generating the eMSTO. \cite{maurya24} observed the same for the open cluster NGC 2355 using Gaia DR3 photometry combined with FLAMES-UVES spectroscopic data.

Despite that, the low-mass regime in the MCs is somewhat unexplored when compared with the high-mass end. Previous studies have shown the same simultaneous correlation between the presence of eMSTO and the cluster mass and between the presence of eMSTO and the cluster age. This degeneracy explains the presence of eMSTO in low-mass clusters by invoking the stellar rotation since those clusters are in the age regime of high rotation activity. Therefore, the low-mass clusters are crucial to understanding the origin of eMSTO. The present work introduces a comprehensive analysis of MSTO widths in a broad sample of MCs star clusters from the VISCACHA\footnote{\url{http://www.astro.iag.usp.br/~viscacha/index.html}} survey \citep{paperI} collected using the SOuthern Astrophysical Research (SOAR) telescope. Thanks to the SOAR Adaptive Module Imager \citep[SAMI;][]{tokovinin16}, we managed to resolve small, crowded regions to observe star clusters in the MCs inside a mass regime lower than the one observed before using the HST. The primary goal is to demonstrate that it is possible to observe the eMSTO phenomenon using VISCACHA ground-based data. Then, by combining deep ground-based photometry with high-resolution space-based data for the method benchmark, we also investigate the dependence of MSTO widths on cluster mass, escape velocity, and age.

This work is organized as follows: Section \ref{sec:data} presents the star cluster data and the differential reddening correction method. Section \ref{sec:method} discusses the parametrization of the eMSTO we employ and the selection of MSTO stars for the MSTO width calculation. Section \ref{sec:discussion} shows the eMSTO in low-mass clusters. Finally, Section \ref{sec:conclusions} draws the conclusions.

%
\section{VISCACHA data}\label{sec:data}

Our star cluster sample selection is based on clusters that showed a more populated CMD from the first VISCACHA internal data release (IDR1): NGC2241, SL28, SL36, SL61, SL576, Kron 37, Lindsay 106, and NGC152. All clusters are within the age range classified as intermediate-age, $\sim 1-3$ Gyr, which predicts the presence of a more evident eMSTO \citep[e.g.][]{yang13,goudfrooij14}.

We exploit the SAMI/SOAR \citep{tokovinin16} instrument, which delivers high-quality images in V and I-bands {(using the Kron-Cousins BVRI set)} due to its adaptive optics module delivering image quality of $\sim 0.3^" - 0.4^"$. The data reduction is based on \cite{diolaiti2000} -- more details can be found in \cite{paperI}. {The adaptive optics module also allows us to reach a photometric limit level $V\sim23$ mag with $\sim 0.1$ mag precision within the clusters' core radius, which cannot be obtained with ground-based telescope, making it possible to analyse stars including in the crowded region of the cluster. }

As a benchmark for the following methodology, we analysed the HST data for the cluster NGC152 in parallel with VISCACHA data. The data were compiled from the literature by \cite{milone23}. We processed the HST data using the same reddening correction and MSTO width measurement methods applied to the VISCACHA data, which are described in the following sections.

\begin{figure}[hbt!]
    \centering
    \includegraphics[width=\columnwidth]{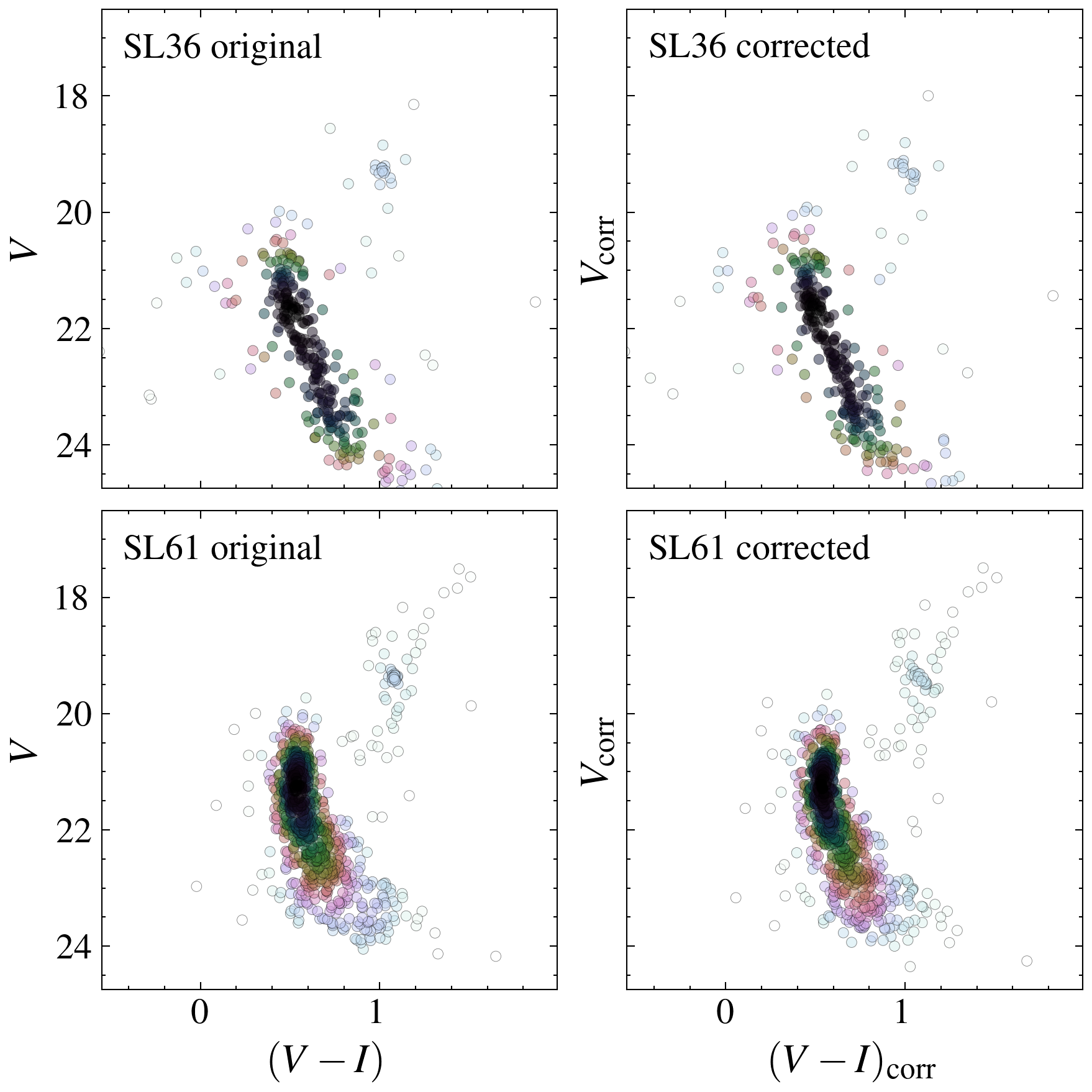}
    \caption{CMDs for the clusters SL36 (top panels) and SL61 (bottom panels) before (left) and after (right) differential reddening correction.}
    \label{fig:sl36_cmd}
\end{figure}

\subsection{Differential reddening correction}

{To account for field star contamination, we applied a statistical decontamination method, as described by \cite{maia10}. This technique compares the CMD of the cluster region (defined as half of the tidal radius, \citealt{ferreira25}) to that of a nearby field region of equal area and similar extinction. The CMDs are divided into cells in the CMD, and stars are statistically removed based on overdensities. Final membership probabilities are assigned based on multiple realisations with varying grid configurations. Only stars with high membership are retained in the analysis, ensuring that field contamination is minimised even in the outer regions of the cluster. We refer the reader to \cite{maia10}, \cite{oliveira23}, \cite{ferreira24} and \cite{ferreira25} for a more detailed description of the method.} 

After that, we adopted the widely used methodology in the literature for the differential reddening correction \citep[e.g.][]{alonsogarcia12,bellini17,souza24}, making some modifications that are briefly described below. The differential reddening correction is particularly important because the spread caused by it can reproduce a fake eMSTO, as observed by \cite{platais12} for the $\sim1.3$ Gyr old open cluster Trumpler 20 using an analysis combining BVI photometry with FLAMES spectroscopic data.

First, a linear transformation is applied to both the CMD and the fiducial line to transform the $V$ vs. $V-I$ CMD in a $V_0$ vs. $E(V-I)$ plane, where $V_0$ and the $E(V-I)$ are the absorption corrected $V$ magnitude and the colour excess, respectively. In this procedure, we used the relative absorption coefficients for $V$ and $I$ bands ($R_V=3.1$ and $R_I=1.87$, from PARSEC database\footnote{https://stev.oapd.inaf.it/cgi-bin/cmd\_3.7}). After the transformation, the distance between each star and the fiducial line in the abscissa axis is the initial guess for their colour excess. A second step is to smooth the initial guess, taking the median colour excess of the k-nearest neighbours in R.A. and DEC. This procedure is repeated until there is no variation on the locus of the fiducial line. After that, the corrected $V_0$ vs. $E(V-I)$ plane is converted into $V$ vs. $V-I$ using the inverse transformation. Then, the difference between the original $V$ and $V-I$ position of the star and its new corrected position is converted into the intrinsic colour excess $E(B-V)$, which is used to construct the differential reddening map.
 
The method is applied typically to stars fainter than the turn-off because the main sequence is more densely populated. However, the clusters analyzed in this work are low mass, then the limited number of stars is a challenge for the traditional reddening correction method. To mitigate this limitation, we implemented a Monte Carlo simulation (MCSim) that takes into account the stellar magnitude uncertainties. The MCSim is used to construct the reddening map across the cluster. The differential reddening corrections for the observed stars are then interpolated using this grid map. Figure \ref{fig:sl36_cmd} presents two examples of the reddening correction: one for a cluster without an eMSTO (SL36, top panels) and another showing evidence of eMSTO (SL61, bottom panels). The differential reddening-corrected CMDs for all clusters are given in Figure \ref{fig:all_cmds}. Even though our method significantly reduces the impact of differential reddening, some residual effects may still be present in the CMDs.

\section{eMSTO Parametrization}\label{sec:method}

\begin{figure*}[hbt!]
    \centering
    \includegraphics[width=0.95\textwidth]{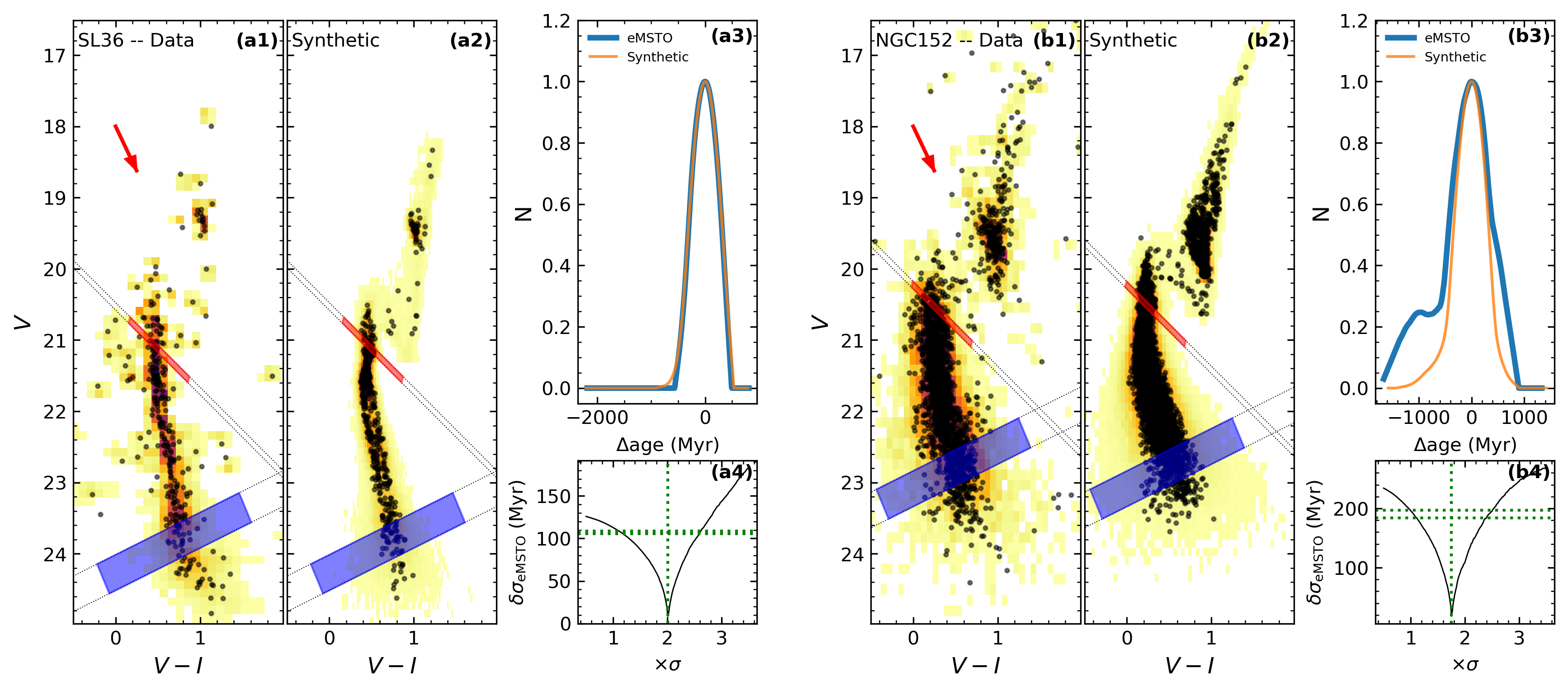}
    \caption{Hess diagrams for SL36 (left) and NGC152 (right) showing the MSTO width derivation. Panels a1 and b1 show the observed data expanded via Monte Carlo simulation. The synthetic CMDs generated from the best isochrone fitting are in panels a2 (SL36) and b2 (NGC152). The red arrow represents the reddening vector for a $E(B-V)=0.2$. The age distribution along the eMSTO are shown in panels a3 (SL36) and b3 (NGC152). The synthetic CMDs were generated assuming the scaled errors using the MS control sample and demonstrated in panels a4 (SL36) and b4 (NGC152). The dotted black lines in panels a1, a2, b1, and b2 are the support lines to construct the eMSTO parallelogram (red shaded region) and the MS control sample (blue shaded region). The vertical green line in panels a4 and b4 shows the best error scale, while the horizontal line is the mean age standard deviation in the MS control sample used as an error for the MSTO width.}
    \label{fig:sl36_res}
\end{figure*}

In previous studies, it has been shown that the intrinsic MSTO spread due to stellar rotation, in theoretical evolutionary models, increases with cluster age up to $\sim 1.5$ Gyr before decreasing again and becoming negligible around 3 Gyr \citep{yang13, niederhofer15}. Also, the observed MSTO width, in star clusters, when expressed in terms of age spread, follows the same trend \citep{goudfrooij14}.  Although the MSTO width in terms of age spread could suggest prolonged star formation, it serves as a useful, standardized measure, allowing for comparisons across different photometric systems. Therefore, in the following discussion, we adopt the MSTO width in the age spread form as an indirect indicator of the presence of an eMSTO.

To obtain a representative sample of the eMSTO stars, we adapted the methodology described in \cite{goudfrooij14} and \cite{correnti21}, summarized in Figure \ref{fig:sl36_res}. The approach involves defining the eMSTO region within a parallelogram (solid lines and red-shaded region in panels a1, a2, b1, and b2 of Figure \ref{fig:sl36_res}), specifically in the area where isochrones are sensitive to age differences.

The left and right boundaries of the parallelogram are set as perpendicular lines to the reddening vector (red arrow in Figure \ref{fig:sl36_res}), enclosing all MSTO stars along the full extent of the color spread. The upper and lower boundaries are approximately aligned with the reddening vector, as differential reddening effects are {reduced} in that direction. The upper boundary is positioned 0.75 magnitudes above the turn-off, corresponding to the turn-off of equal-mass binaries. The lower boundary varies for each cluster to ensure that at least 1\% of the stars are included within the final parallelogram. Panels a1, a2, b1, and b2 of Figure \ref{fig:sl36_res} illustrate the final parallelogram for SL36 (left) and NGC152 (right) as a solid black outline with the red-shaded region.

The stars within the parallelogram are then interpolated into a grid of isochrones with different ages to derive their age values. Our isochrone grid is based on the PARSEC database \citep{bressan12}, covering ages from 700 Myr to 3.5 Gyr in 100 Myr steps. We also account for a range of metallicities\footnote{PARSEC isochrones define metallicity as the total metal-to-hydrogen ratio, [M/H].}, spanning from $-1.0$ to $0.3$ with a fine step size of 0.01. The resulting age distribution is then smoothed using an \texttt{Epanechnikov} kernel, which is less sensitive to standard deviation compared with a Gaussian kernel (see panels a3 and b3 of Figure \ref{fig:sl36_res}).



Next, we constructed the synthetic single stellar population (SSP) cluster using the cluster parameters obtained from the best isochrone fit of the cluster using the SIESTA\footnote{Publicly available at \url{https://github.com/Bereira/SIESTA/}} code \citep{ferreira24}. Cluster mass values were taken from \cite{santos20}, except for NGC152, for which we adopted the estimate from \cite{gatto21}. A summary of all adopted values and references is provided in Table \ref{tab:my_label}. We also employed the \cite{kroupa01} initial mass function and excluded synthetic stars to replicate the observed cluster luminosity function. Notably, our synthetic population was designed to closely reproduce the observed CMD, including the presence of binary stars. 

To account for possible under- or overestimated magnitude errors, we also defined a control parallelogram located at of the main sequence few magnitudes fainter than the turn-off \citep[blue shaded region in panels a1, a2, b1, and b2 of Figure \ref{fig:sl36_res}, following][]{correnti21}. This control region helps assess the impact of photometric errors on the synthetic population. We systematically increased the contribution of errors in the synthetic CMD until the standard deviation in the control region matched that of both the observed and synthetic CMDs. The remaining residual difference was then considered as an additional source of uncertainty in our MSTO width determinations (panels a4 and b4 of Figure \ref{fig:sl36_res}).  

Finally, similar to the differential reddening correction, we found that the MCSim improved the accuracy of our calculations. Therefore, we applied the parallelogram selection and MSTO width determination within the MCSim framework.  

\begin{figure*}[hbt!]
    \centering
    \includegraphics[width=0.99\textwidth]{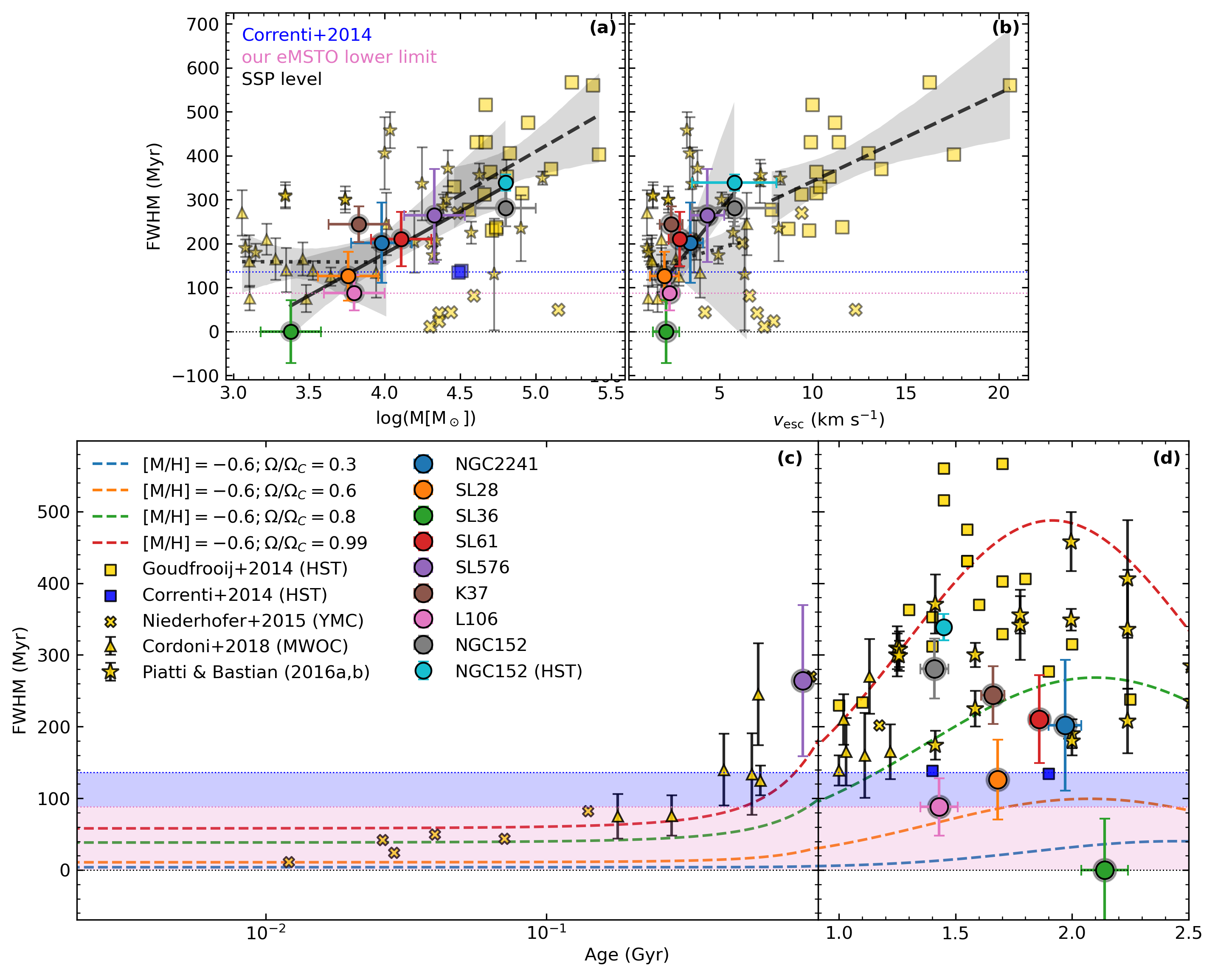}
    \caption{The MSTO width at 50\% of the maximum (FWHM), as a function of cluster mass (panel a), escape velocity (panel b), and age (panels c and d). The colored dots represent the star clusters analyzed in this work. Yellow squares represent MCs clusters from \cite{goudfrooij14}, the blue squares are the clusters analyzed by \cite{correnti14} and used by \cite{goudfrooij14} lower limit for the existence of the eMSTO, yellow crosses indicate YMC analyzed by \cite{niederhofer15}. Yellow triangles show MWOC from \cite{cordoni18}. The yellow stars are the results by \cite{piatti16mnras,piatti16aea}. The colored dotted lines illustrate the expected MSTO width comparing non-rotating isochrones to models for different rotation fractions $\Omega/\Omega_C=$ 0.3 (blue), 0.6 (orange), 0.8 (green), and 0.99 (red), adopting the metallicity of [M/H]$=-0.6$ (mean of our sample). The horizontal dashed blue line represents previous limits by \cite{correnti14} and \cite{goudfrooij14}, while the pink line and shaded region mark the eMSTO limit derived in this work. The black lines in panels (a) and (b) represent the linear regression for the data from \citet[][dotted]{cordoni18}, \citet[][dashed]{goudfrooij14}, and this work (solid). The gray shaded regions represent the standard deviation of the fit. }
    \label{fig:main_fig}
\end{figure*}

After satisfying the control test in the MS, the age distributions for the original MCSim and synthetic CMD eMSTO are obtained. The width is calculated as follows \citep{goudfrooij14}:
\begin{equation}
    {\rm FWHM} = \sqrt{ {\rm FWHM}_{\rm cluster}^{2} - {\rm FWHM}_{\rm synthetic}^{2} },
\end{equation}
where FWHM is the full width at half of the maximum. The values derived for all clusters in this work are summarised in Table \ref{tab:my_label} and shown in Figure \ref{fig:main_fig}.

\begin{table*}
\caption{Fundamental parameters assumed for this work (3 to 10) and values derived from our analysis (13 and 14). }
\centering
\resizebox{0.99\linewidth}{!}{
\begin{tabular}{lcccccccccccccc}

\hline
\hline
\noalign{\smallskip}
 & \multicolumn{4}{c}{Input/Literature Values} & & \multicolumn{6}{c}{Isochrone fitting} & & \multicolumn{2}{c}{This work} \\
\noalign{\smallskip}
\cline{2-5}  \cline{7-12} \cline{14-15} 
\noalign{\smallskip}
Cluster & MC & $r_h$ & $\log{\rm Mass}$ & $v_{esc}$  & & Age & [Fe/H] & (m$-$M)$_0$ & $E(B-V)$ & $f_{\rm bin}$ & REF & & FWHM & $\times\sigma$\\  
\noalign{\smallskip}

        &  & pc & $\log{[M_\odot]}$ & km$\,$s$^{-1}$  & & Gyr &        &             &          &            \% & & & Myr  &                \\
\noalign{\smallskip}
 (1) & (2) & (3) & (4) & (5) & & (6) & (7) & (8) & (9) & (10) & (11) & & (12) & (13) \\
\noalign{\smallskip}
\hline
\noalign{\smallskip}

      NGC~2241 & LMC & $4.8\pm1.0$ & $3.98\pm0.14$ & $3.46\pm0.69$ & & $1.97\pm0.07$  &    $-0.51\pm0.06$    &    $18.36\pm0.04$  &     $0.11\pm0.01$     &    $0.54\pm0.08$  & FIP & &   $202\pm91$     & 3.5  \\
\noalign{\smallskip}
      SL28     & LMC & $8.3\pm2.7$ & $3.76\pm0.14$ & $2.00\pm0.62$ & & $1.68\pm0.03$  &    $-0.43\pm0.03$    &    $18.49\pm0.02$  &     $0.12\pm0.01$     &    $0.47\pm0.03$  & FIP & &   $126\pm56$     & 2.4  \\
\noalign{\smallskip}
      SL36     & LMC & $3.0\pm0.4$ & $3.38\pm0.26$ & $2.16\pm0.70$ & & $2.14\pm0.10$  &    $-0.65\pm0.09$    &    $18.59\pm0.06$  &     $0.11\pm0.01$     &    $0.40\pm0.09$  & FIP & &   $  0\pm71$     & 2.0 \\
\noalign{\smallskip}
      SL61     & LMC & $9.2\pm2.1$ & $4.11\pm0.14$ & $2.82\pm0.62$ & & $1.86\pm0.04$  &    $-0.33\pm0.05$    &    $18.53\pm0.03$  &     $0.11\pm0.01$     &    $0.49\pm0.06$  & FIP & &   $211\pm62$     & 2.6  \\
\noalign{\smallskip}
      SL576    & LMC & $6.6\pm1.3$ & $4.33\pm0.16$ & $4.43\pm0.95$ & & $0.82\pm0.04$  &    $-0.20\pm0.07$    &    $18.13\pm0.06$  &     $0.13\pm0.01$     &    $0.39\pm0.06$  & FIP & &   $264\pm105$     & 3.5 \\
\noalign{\smallskip}
      K37      & SMC & $7.0\pm2.0$ & $3.83\pm0.14$ & $2.37\pm0.64$ & & $1.58\pm0.01$  &    $-0.71\pm0.01$    &    $19.17\pm0.02$  &     $0.01\pm0.03$     &    $0.59\pm0.11$  & SIP & &   $244\pm40$     & 1.9  \\
\noalign{\smallskip}
      L106     & SMC & $6.9\pm1.1$ & $3.80\pm0.16$ & $2.34\pm0.47$ & & $1.53\pm0.02$  &    $-0.81\pm0.05$    &    $18.48\pm0.08$  &     $0.07\pm0.02$     &    $0.33\pm0.12$  & SIP & &   $88\pm40$     & 1.6  \\
\noalign{\smallskip}
     \hline
\noalign{\smallskip}
      NGC~152                  & SMC & $11.2\pm2.1\,^*$ & $4.80\pm0.30\,^*$ & $5.66\pm2.21$ & & $1.41\pm0.06$  &    $-0.85\pm0.03$    &    $18.89\pm0.07$  &     $0.11\pm0.02$     &    $0.38\pm0.10$  & PXI & &  $281\pm42$     & 1.8  \\
\noalign{\smallskip}
      NGC~152 (HST)$^\dagger$  & SMC & $11.2\pm2.1\,^*$ & $4.80\pm0.30\,^*$ & $5.66\pm2.21$ & & $1.45\pm0.06$  &    $-0.60\pm0.03$    &    $19.07\pm0.07$  &     $0.03\pm0.02$     &    $0.10\pm0.10$  & M23 & & $339\pm19$     & 1.6  \\
\noalign{\smallskip}
\noalign{\smallskip} \hline 
\end{tabular}}
\label{tab:my_label}
\tablebib{FIP (Ferreira in prep); SIP (Saroon in prep); PXI \citep{ferreira24}; M23 \citep{milone23}. 
Note: $^\dagger$ Data and fundamental parameters taken from \cite{milone23}, while the uncertainties were assumed the same as to VISCACHA data. * Values taken from \cite{gatto21}.}
\end{table*}



\section{The eMSTO in low-mass clusters}\label{sec:discussion}

To benchmark our method, we applied the same analysis to NGC152 using both HST (cyan dot in Figure \ref{fig:main_fig}) and VISCACHA (grey dot in Figure \ref{fig:main_fig}) data. Further details can be found in Appendix \ref{app:benchmark} and Figure \ref{fig:ngc152_hst}. The MSTO width values obtained from both datasets agree between $1\sigma-2\sigma$. Since our analysis helps mitigate the effects of differential reddening and photometric errors, the remaining differences could be attributed to variations in the isochrone fitting values between the VISCACHA and HST datasets. \cite{milone23} selected the isochrone that best fits the upper density of the MSTO as the optimal fit. Despite this difference in approach, our validation test confirms that the methodology is consistent across different photometric systems. Furthermore, the agreement in MSTO width between both photometric datasets reinforces that our method effectively identifies the presence of eMSTO, which is clearly visible in HST data. Another key observation is the existence of a cluster without eMSTO (green dot in Figure \ref{fig:main_fig}), which is represented by a null width. This indicates that its observed width matches that of a synthetic SSP, confirming the absence of eMSTO in this case.

We performed the generalised Kendall correlation test to verify the correlation between the MSTO width and the parameters involved in this work. All the values are summarised in Table \ref{tab:kendall}. We compare with the values taken directly from \cite{goudfrooij14}. The cluster mass and escape velocity values assumed in this work from \cite{goudfrooij14} are regarding the present day to have a direct comparison to our results.

The linear relation between the MSTO width and the cluster mass (in log scale, panel a of Figure \ref{fig:main_fig}) and escape velocity (panel b) reported by \citet[][yellow squares]{goudfrooij14} is well extended by our results in the low-mass regime (coloured dots). The (present-day) escape velocity was derived using the same equation as \cite{georgiev09} and \cite{cabrera-ziri16a}:
\begin{equation}
    \displaystyle v_{esc} = f_c \times \sqrt{\frac{M_{cls}[M\odot]}{r_{\rm eff}[pc]}} \,\,\,[km\,s^{-1}],
\end{equation}
where $r_{\rm eff}$ is the effective radius. For the parameter $f_c$, we assume the same value as \cite{cabrera-ziri16a} of $0.076$ that was calculated for diffuse clusters using a \cite{king62} profile given by \cite{georgiev09}. We assume the half-light radius as the effective radius \citep[as in][]{georgiev09} from \cite{santos20} except for NGC152, which we assumed the value from \cite{gatto21}. The Kendall correlation test returned a strong positive correlation between both mass and escape velocity with the MSTO width represented by the positive $\tau$ value and low $p_\tau$ value. Besides that, the Kendall test values are similar to those obtained by \cite{goudfrooij14}, also observed by the linear regression in panels a and b of Figure \ref{fig:main_fig}.

Since the cluster masses adopted in this work refer to their present-day values, one possible explanation for having similar MSTO widths across clusters with different current masses is the effect of tidal interactions experienced within the dwarf galaxy environment \citep[e.g.][]{moreno-hilario24}. Thus, while the initial mass may be key to the origin of the eMSTO, the current mass, when considered alongside the MSTO width, can also offer valuable insights into the cluster’s dynamic history and the broader galactic environment of the MCs.

Although our sample displays a clear linear correlation between the MSTO width and the cluster mass/$v_{esc}$, the low-mass clusters analysed by \citet[][yellow stars in Figure \ref{fig:main_fig}]{piatti16mnras,piatti16aea} do not present any trend. In their sample, two clusters have MSTO width $\sim300$ Myr while the other two $\sim200$ Myr. Nevertheless, when observing their ages (panels c and d), the two clusters with high MSTO width are in perfect agreement with the pattern observed for \cite{goudfrooij14} data and explained by stellar rotation, as claimed by \cite{piatti16mnras,piatti16aea}. On the other hand, the two clusters with low MSTO width follow the same behaviour as our values for NGC2241 (blue) and SL61 (red), showing a possible common stellar rotation rate sequence. For completeness purposes, we also included the MWOCs from \citet[][yellow triangles in Figure \ref{fig:main_fig}]{cordoni18}, which indicates a plateau around $150$ Myr.

In comparison with \citet[][yellow crosses in Figure \ref{fig:main_fig}]{niederhofer15} for YMCs of MCs using HST, our sample also shows essentially higher values of MSTO width. The three clusters in our sample that have width values below the limit represented by the blue squares and the blue dashed line in Figure \ref{fig:main_fig} agree with the \cite{niederhofer15} sample. Looking at panels c and d of Figure \ref{fig:main_fig}, it is clear that the relationship between the width and the cluster mass (and escape velocity) also depends on the cluster age \citep{yang13, niederhofer15}. It suggests that the eMSTO would be a transient phenomenon in terms of age but intrinsically related to the cluster mass. From the high Kendall $p_\tau$ values, no correlation is found between the MSTO width and the cluster age.

{Although the eMSTO feature was first identified in star clusters from the MCs, it has also been observed in Milky Way open clusters \citep[MWOCs,][]{marino18}. This finding is significant because MWOCs can be similarly young ($<2$ Gyr), suggesting that eMSTOs may be a common characteristic of young star clusters, regardless of their environment. In MWOCs, \citet{marino18} also detected a substantial population of fast-rotating stars in the MSTO region, supporting the hypothesis that stellar rotation plays a central role in the eMSTO phenomenon. Moreover, this mechanism could represent the early phase of the multiple stellar population phenomenon observed in older globular clusters. In this context, fast-rotating stars in young clusters could enrich the intra-cluster medium, leading to the chemical variations later seen in clusters older than $\sim2$ Gyr \citep{martocchia18} and globular clusters \citep{bastian2018,milone22}. The young massive clusters from \citet{niederhofer15} do not yet exhibit eMSTOs in their CMDs, likely because their ages are still below the threshold where rotation-induced effects become observable. In contrast, our sample covers an age range associated with peak stellar rotation efficiency, allowing us to detect a clear correlation between MSTO width and cluster mass. This age-dependent trend is also present in the MWOCs studied by \citet{cordoni18}, which match the younger end of the \citet{niederhofer15} sample (see panel c of Figure \ref{fig:main_fig}).}

To explain the large width spread in panel d of Figure \ref{fig:main_fig}, we compared our results with stellar rotation models using PARSEC isochrones \citep{nguyen22} with a fixed metallicity of [M/H] = -0.60, approximately the mean metallicity of our sample. {The theoretical MSTO widths were derived using the same approach as for the observed data. We first select the isochrone with a specific rotation rate. Then, we consider all the available inclination angles. With the set of isochrones having the same metallicity and rotation rate but different inclination angles, we interpolate the points as made for the cluster CMDs. The reference grid is composed of non-rotating isochrones of different ages. Finally, to handle any offset, we use the isochrone with an inclination angle of 60 degrees, which perfectly reproduces the isochrone without rotation. We repeat this procedure for all the ages from 10 Myr to 3 Gyr.}. We find that clusters with broader eMSTOs align with higher rotation rates ($\Omega/\Omega_C \geq 0.5$), supporting the scenario where differential rotation broadens the MSTO. 


Our result complements the lack of low-mass clusters in the study of the eMSTO. The presence of eMSTOs in low-mass clusters indicates a more universal mechanism, potentially linked to fundamental stellar evolutionary processes rather than specific cluster properties. Future investigations, including spectroscopic follow-ups and more detailed modelling, will be crucial to disentangling the relative contributions of rotation, age spreads, and cluster dynamics in shaping the eMSTO phenomenon.


\begin{table}
\caption{Correlation coefficient and p-value for a generalized Kendell test.}
\centering
\resizebox{0.95\columnwidth}{!}{
\begin{tabular}{rcccc}

\hline
\hline
\noalign{\smallskip}
\noalign{\smallskip}
 & \multicolumn{2}{c}{This work} &  \multicolumn{2}{c}{\cite{goudfrooij14}}  \\
\noalign{\smallskip}
Combination & $\tau$ & $p_\tau$ & $\tau$ & $p_\tau$ \\  
\noalign{\smallskip}
 (1) & (2) & (3) & (4) & (5)   \\
\noalign{\smallskip}
\hline
\noalign{\smallskip}

FWHM -- logMass     & 0.7857 & 0.0055 & 0.8000 & 0.0132 \\
\noalign{\smallskip}
FWHM -- $v_{esc}$   & 0.6429 & 0.0312 & 0.9684 & 0.0027 \\
\noalign{\smallskip}
FWHM -- age         & -0.5000 & 0.1087 & 0.0067 & 0.9696  \\

\noalign{\smallskip}
\noalign{\smallskip} \hline 
\end{tabular} }
\label{tab:kendall}
\end{table}

\section{Conclusions}\label{sec:conclusions}

In this work, we have investigated the presence of extended main-sequence turn-offs (eMSTOs) in a sample of eight intermediate-age star clusters from the VISCACHA survey using deep ground-based photometry from the SOAR telescope. Our analysis explored the dependence of MSTO widths on cluster mass, escape velocity, and age, expanding previous studies to lower-mass cluster end. The main conclusions of our study are summarized as follows:

\begin{itemize}

    \item The comparison between the MSTO widths derived from VISCACHA data and the HST benchmark for NGC152 establishes the robustness of our methodology, validating the reliability of ground-based observations for studying eMSTOs in low-mass clusters.

    \item We detect eMSTO, via non-negligible MSTO width values, in the majority of the analyzed clusters, confirming that the phenomenon is not restricted to high-mass clusters. 
    
    \item One cluster shows no eMSTO represented as a zero MSTO width. However, this cluster is also the oldest in our sample, located in the age regime for which the stellar rotation becomes less efficient. 
    
   \item We find a cluster, L106, with an age at the peak of stellar rotation activity, presents an MSTO width lower than the assumed lower limit before. Therefore, we propose a new lower limit for the presence of eMSTO around 90 Myr for the MSTO width.    

    \item Our results reveal a correlation between MSTO width and cluster mass, as well as escape velocity, consistent with previous findings \citep[e.g.,][]{goudfrooij14, correnti14}.
    
    \item The MSTO widths in our sample also exhibit the predicted age dependence, peaking at intermediate ages and decreasing at older ages, consistent with the scenario where stellar rotation effects maximize near the end of the main sequence and diminish due to angular momentum loss \citep[e.g.,][]{georgy19} confirming a phenomenon known as gravity darkening.
    
\end{itemize}

These results expand our understanding of eMSTOs in star clusters, particularly in the low-mass regime. They suggest that stellar rotation could be a dominant factor influencing eMSTO morphology, but not solely, as previous studies have suggested. Our results show the reliability of using ground-based photometry and the power of the adaptive optics instrument at SOAR once more. The VISCACHA collaboration has a great opportunity to study the eMSTO origin in more detail for the low-mass end since we possess observations for star clusters spread along the Small and Large MCs as well as in the Magellanic Bridge. Further investigations also combining spectroscopic data to probe rotational velocities and chemical abundances will be crucial to fully unravel the origin of eMSTO in star clusters.

%
\begin{acknowledgements}
{The authors are grateful to the anonymous referee for his/her suggestions and remarks, which greatly improved the paper}. Many thanks to Dr. Thanh Nguyen for the discussion and insights about the new set of PARSEC isochrones. We thank Peter Smith from the Galactic Nuclei group at MPIA for the discussion about escape velocity. SOS acknowledges the support from Dr. Nadine Neumayer's Lise Meitner grant from the Max Planck Society, and FAPESP (proc. 2018/22044-3). APV and SOS acknowledge the DGAPA–PAPIIT grant IA103224. BD acknowledges support by ANID-FONDECYT iniciación grant No. 11221366 and from the ANID BASAL project FB210003. L.O.K. acknowledges partial financial support by CNPq (proc. 313843/2021-0) and UESC (proc. 073.6766.2019.0013905-48). BPLF acknowledges financial support from Conselho Nacional de Desenvolvimento Científico e Tecnológico (CNPq, Brazil; proc. 140642/2021-8) and Coordenação de Aperfeiçoamento de Pessoal de Nível Superior (CAPES, Brazil; Finance Code 001; proc. 88887.935756/2024-00). BB acknowledges partial financial support
from FAPESP, CNPq and CAPES financial code 001. E.R.G. gratefully acknowledges support from ANID PhD scholarship No. 21210330. GB acknowledges CONICET PIP 112-202101-00714 and Agencia I+D+i PICT 2019-0344. D.M. gratefully acknowledges support from the ANID BASAL projects ACE210002 and FB210003, from Fondecyt Project No. 1220724, and from CNPq Brasil Project 350104/2022-0. FFSM acknowledges support from CNPq (proc. 404482/2021-0) and FAPERJ (proc. E-26/201.386/2022 and E-26/211.475/2021). J.G.F-T gratefully acknowledges the grant support provided by Proyecto Fondecyt Iniciaci\'on No. 11220340, and also from the Joint Committee ESO-Government of Chile 2021 (ORP 023/2021). E.B. acknowledges  CNPq for support. NSF–DOE Vera C. Rubin Observatory is a Federal project jointly funded by the National Science Foundation (NSF) and the Department of Energy (DOE) Office of Science, with early construction funding received from private donations through the LSST Corporation. The NSF-funded LSST (now Rubin Observatory) Project Office for construction was established as an operating center under the management of the Association of Universities for Research in Astronomy (AURA). The DOE-funded effort to build the Rubin Observatory LSST Camera (LSSTCam) is managed by SLAC National Accelerator Laboratory (SLAC). The work is based on observations obtained at the Southern Astrophysical Research (SOAR) telescope (projects SO2016B-018, SO2017B-014, CN2018B-012, SO2019B-019, SO2020B-019, SO2021B-017), which is a joint project of the Ministério da Ciência, Tecnologia, e Inovação (MCTI) da República Federativa do Brasil, the U.S. National Optical Astronomy Observatory (NOAO), the University of North Carolina at Chapel Hill (UNC) and Michigan State University (MSU).
\end{acknowledgements}

\bibliography{bibliography}{}
\bibliographystyle{aa}

\appendix

\counterwithin{figure}{section}
%


\section{Corrected CMD of all clusters}\label{app:cmds}
\begin{figure*}
    \centering
    \includegraphics[width=\textwidth]{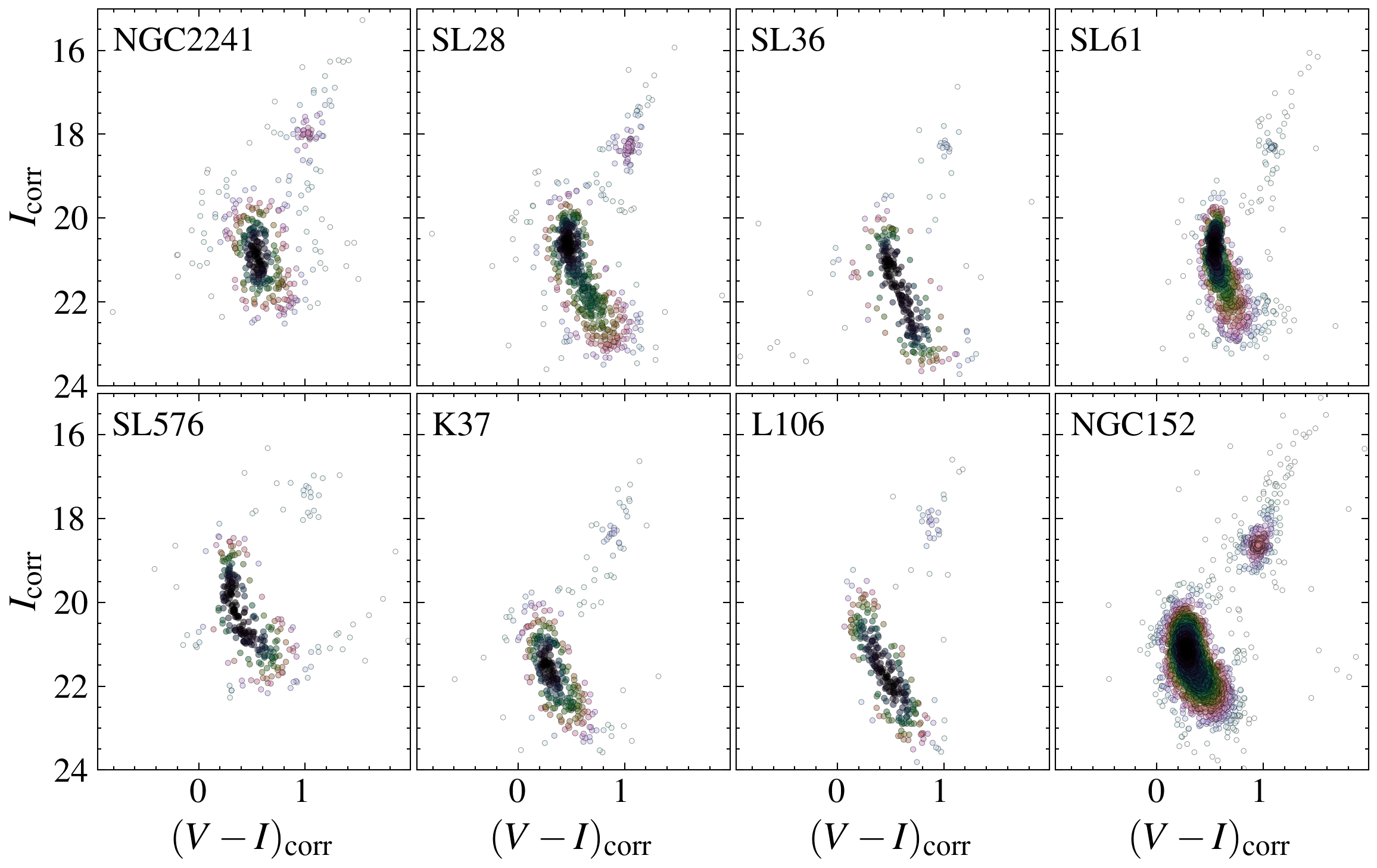}
    \caption{CMD corrected by differential reddening of all clusters in our sample.}
    \label{fig:all_cmds}
\end{figure*}

Figure \ref{fig:all_cmds} shows the differential reddening corrected CMD for all clusters analyzed in this work.

\section{Benchmark -- NGC152 HST Data}\label{app:benchmark}

To verify the feasibility of our data, we analyzed the cluster NGC 152, which was also observed by the HST, providing an ideal benchmark for comparison. NGC 152 is a well-studied intermediate-age cluster in the Small Magellanic Cloud. It is known for its pronounced extended main-sequence turn-off (eMSTO), which has been widely investigated to understand age spreads and stellar rotation effects in star clusters.  

We performed the decontamination of field stars using the method presented by \cite{gallart03} and described by \cite{marino14,milone23}, which is essentially the same method as by \cite{maia10} widely used in the VISCACHA studies. These approaches account for the statistical subtraction of field star contamination by considering the spatial distribution and photometric properties of stars in the cluster's vicinity. We used data from \cite{milone23}, which provides a compilation from the HST archival. 

To correct for differential reddening, we adopted the same procedure used for the VISCACHA data, which is explained in the main text. This step is crucial for accurately measuring the MSTO width, as differential reddening can artificially broaden the main-sequence turn-off region.

\begin{figure}
    \centering
    \includegraphics[width=\columnwidth]{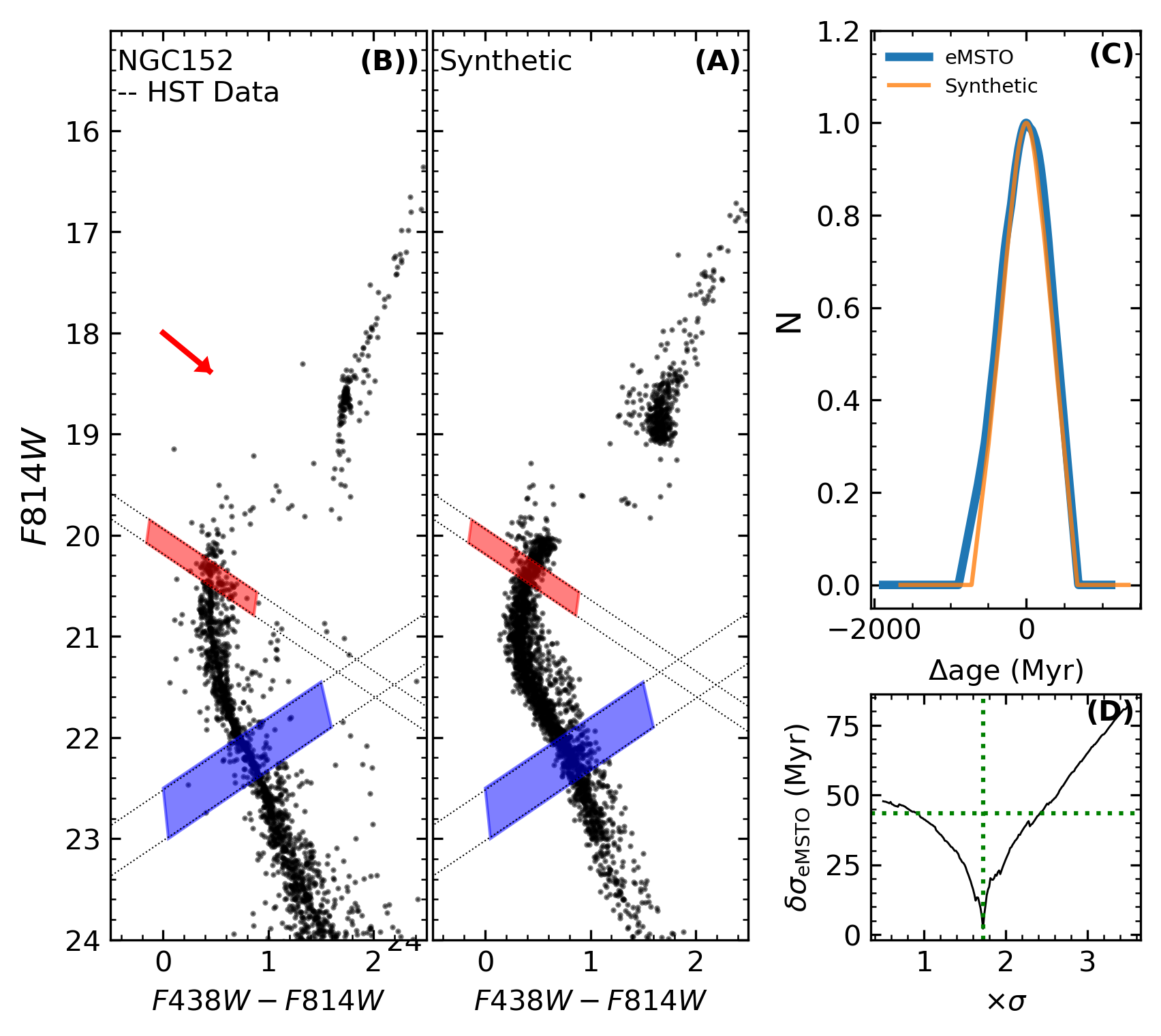}
    \caption{Same as Figure \ref{fig:sl36_res}, but for NGC 152 using HST data.}
    \label{fig:ngc152_hst}
\end{figure}

After applying these corrections, we measured the MSTO widths from both the VISCACHA and HST data (Figure \ref{fig:ngc152_hst}). The results are consistent within $1.3\sigma$, indicating good agreement between the two datasets. This consistency validates the reliability of our decontamination and differential reddening correction methods, supporting the robustness of our analysis.

\end{document}